\documentclass[journal=nalefd,manuscript=article]{achemso}

\usepackage{amsmath}
\usepackage{amssymb}

\usepackage{chemformula} 

\newcommand{\inse}{In${}_2$Se${}_3$ }

\newcommand\hl[1]{%
  \bgroup
  \hskip0pt\color{black!80!black}%
  #1%
  \egroup}

\author{Jean Spièce}
\affiliation{\small \textit Institute of Condensed Matter and Nanosciences, Université catholique de Louvain (UCLouvain), 1348 Louvain-la-Neuve, Belgium}
\email{jean.spiece@uclouvain.be, pascal.gehring@uclouvain.be }
\author{Valentin Fonck}
\affiliation{\small \textit Institute of Condensed Matter and Nanosciences, Université catholique de Louvain (UCLouvain), 1348 Louvain-la-Neuve, Belgium}
\author{Charalambos Evangeli}
\affiliation{\small \textit Institute of Condensed Matter and Nanosciences, Université catholique de Louvain (UCLouvain), 1348 Louvain-la-Neuve, Belgium}
\author{Phil S. Dobson}
\affiliation{\small \textit James Watt School of Engineering, University of Glasgow,  Glasgow, G12 8LT, United Kingdom}
\author{Jonathan M. R. Weaver}
\affiliation{\small \textit James Watt School of Engineering, University of Glasgow,  Glasgow, G12 8LT, United Kingdom}
\author{Pascal Gehring*}
\affiliation{\small \textit Institute of Condensed Matter and Nanosciences, Université catholique de Louvain (UCLouvain), 1348 Louvain-la-Neuve, Belgium}

\title{Direct measurement of the local electrocaloric effect in 2D ferroelectric In${}_2$Se${}_3$ by Scanning Electrocaloric Thermometry}

\begin{document}

\maketitle



\begin{abstract}
The electrocaloric effect refers to the temperature change in a material when an electric field is applied or removed. Significant breakthroughs revealed its potential for solid-state cooling technologies in past decades. These devices offer a sustainable alternative to traditional vapor compression refrigeration, with advantages such as compactness, silent operation, and the absence of moving parts or refrigerants.

Electrocaloric effects are typically studied using indirect methods using polarization data, and which suffer from inaccuracies related to assumptions about heat capacity. Direct methods, although more precise, require device fabrication and face challenges in studying meso- or nanoscale systems, like 2D materials, and materials with non-uniform polarization textures where high spatial resolution is required.

In this study, a novel technique, Scanning Electrocaloric Thermometry, is introduced for characterizing the local electrocaloric effect in nanomaterials. This approach achieves high spatial resolution by locally applying electric fields and by simultaneously measuring the resulting temperature change. By employing AC excitation, the measurement sensitivity is further enhanced and the electrocaloric effect is disentangled from other heating mechanisms such as Joule heating and dielectric losses. The effectiveness of the method is demonstrated by examining electrocaloric and heat dissipation phenomena in two-dimensional \inse micrometer-sized flakes.
\end{abstract}

\section{Introduction}
The electrocaloric effect (ECE) describes a temperature change in a material when an electric field is applied or removed. The dipoles in a dielectric material will align in the presence of an electric field, therefore reducing the configurational entropy of the system and thus, increasing its temperature (see \textbf{Figure \ref{fig1}a}). Although, the effect was first predicted by William Thomson\cite{thomson1878ii} in 1878, it was not observed experimentally before the mid-60 in lead zirconate (PZT). A seminal study revived the research community's interest in 2006 when a significant ECE was measured in PZT thin films\cite{mischenko2006giant}, opening up new strategies for solid-state cooling technologies.

From quantum technologies to air-conditioning through vaccines conservation, solid-state cooling devices represent an attractive prospect in the field of refrigeration. Their main advantage lies in their simplicity and the absence of mechanical components or refrigerants, which makes them compact, silent, extremely reliable and sustainable compared to vapor compression methods and thermoelectric materials. Furthermore, caloric heat pumping cycle have been shown to have high coefficient of performance compared to other methods. 

The ECE is usually studied in polarizable materials, such as para- and ferroelectric materials. In those materials, electric dipoles align with the external electric field leading to an effective polarization $P$. As the electric field is varied, the polarization varies and the dipole configuration entropy $S$ increases or reduces. The entropy reduction, when the dipoles align with the electric field, is an exothermic process, leading to a temperature increase $\delta T_\mathrm{ECE}$ in the material: 
\begin{equation}
    \delta T_\mathrm{ECE} \propto \frac{\delta S}{\delta t} \propto \frac{\delta P}{\delta t}.
\end{equation}
In most materials, especially in ferroelectrics, the polarization is not linearly proportional to the electric field strength. This leads to non-linearities in the materials temperature change under different electric field. 

When trying to identify novel materials with high electrocaloric performance and to understand materials parameters and microscopic mechanisms responsible for such high performance, two approaches have been widely reported. On the one hand, most studies use indirect methods that rely on the measurements of the field- and temperature-dependent polarization $P(E,T)$. Using Maxwell's relations, the adiabatic temperature change can be written as 
\begin{equation}
    \Delta T_\mathrm{ECE}^\mathrm{ad} = - \int_0^{E_\mathrm{max}} C_V^{-1} T \left( \frac{ \partial P}{\partial T} \right)_E \mathrm{d}E, 
\end{equation}
where $C_V$ and $E_\mathrm{max}$ are the volumetric heat capacity and maximum applied electric field, respectively. Indirect methods suffer from the hypothesis that $C_V$ does not depend on $E$ leading to errors in the $\Delta T_\mathrm{ECE}^\mathrm{ad}$ measurement. 
Direct methods, on the other hands, do not suffer from those problems, as the temperature variation is directly measured on the sample when applying or removing an electric field, but measuring temperatures accurately, in particular in small samples is challenging. The most common direct method that does not require the implementation of complicated microscopic thermometers in a device uses infrared thermography and its lock-in variation\cite{liu2016direct,kar2013direct,fischer2023simultaneous,pandya2017direct,iguchi2023direct}. However, since this method lacks high spatial resolution, it can neither be applied to mesoscale and nanoscale systems (like 2D materials, nano-wires, nano-patterned structures, ...) nor to study materials with nonuniform polarization texture or nano-regions domain structure.

Here, we introduce a novel method for characterizing the local electrocaloric effect in nanomaterials without the need of device fabrication. Our proposed approach combines high spatial resolution and direct measurement of the temperature using a scanning thermal microscope (SThM). SThM has been used for electrocaloric measurements of bulk sample already\cite{baxter2023high,liao2024revealing} without benefiting from the high spatial resolution of the SThM tip. We achieve true locality by using the SThM probe directly to locally apply electric fields, to drive a local electrocaloric effect and to then locally measure the resulting temperature change with the same SThM probe (see Figure \ref{fig1}b). Furthermore, while most direct measurements schemes use DC pulses to measure the temperature rise and fall within a material, we use AC excitation\cite{pandya2017direct,iguchi2023direct} which enables high-sensitivity and the distinction between entangled effects such as the electrocaloric and Joule heating or dielectric losses. We demonstrate the excellent spatial and temperature resolution of our method by characterizing the heat dissipation and electrocaloric effects in 2-dimensional In$_2$Se$_3$ $\mu$m-sized flakes.

\section{Results and Discussion}

\subsection{Characterization Method and sample preparation}

\begin{figure}[h]
  \includegraphics[width=\linewidth]{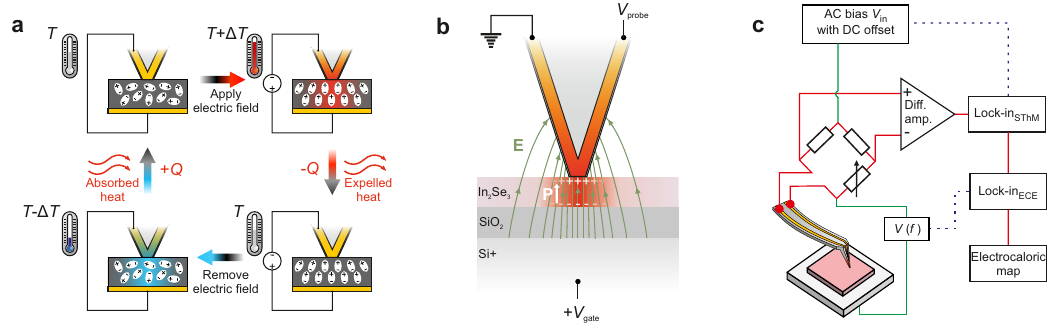}
  \caption{\textbf{Electrocaloric and thermometry principles}. (a) Thermal cycle of the ECE effect in the material with the SThM probe used for applying an electric field and for the temperature measurement. (b) Schematic of the electric field line distribution when approaching a sample (with a voltage $V_\mathrm{g}$ on the Si back gate) with a metallic probe. The polarization in the material is also shown. (c) AC thermometry method employed. The signal demodulated at 91kHz in the lock-in labelled SThM is fed into the ECE lock-in for the second demodulation step which gives the final ECE map.}
  \label{fig1}
\end{figure}

Applying an electrical field to a dielectric material can trigger various effects that lead to temperature variations. The power $P_C$ is dissipated in the dielectric with 
\begin{equation}
    P_C = f C U^2 \tan \delta,
    \label{PowC}
\end{equation}
where $f$ and $U$ are the frequency and amplitude of the applied AC voltage, $C$ is the capacitance of the system and $\delta$ is the loss angle of the dissipated electrical power. If there is a leakage current flowing through the dielectric with resistance $R$, further power due to the Joule effect is dissipated with $P_\mathrm{Joule} = U^2/R $. The electric field can also align the material's dipoles and entropy is then reduced, which is countered by a temperature rise via the electrocaloric effect (ECE). Then, when the field is removed, entropy increases back as dipoles start misaligning and the material temperature then decreases. In Figure \ref{fig1}(a), a scheme of the effect under an SThM tip is shown.

We now explore electric field induced local cooling/heating effects in thick layer of \inse.  We apply an AC signal with frequency $f$ on the highly doped Si back gate electrode leading to a modulation of the electric field on the tip-sample system as: 
\begin{equation}
    \text{Bipolar :} \quad E(t) \propto V_0 sin(\omega t) \quad \text{Unipolar :} \quad E(t) \propto \frac{V_0}{2} (1 + sin(\omega t))
\end{equation}
where $\omega = 2 \pi f$. As we shown later, we compute the field strength using realistic finite element models.

To detect electrostatic-related temperature change in our sample, we used a slightly modified Scanning Thermal Microscope (SThM) configuration (see Figure \ref{fig1}(b)). Using the deflection of a laser on the back on the probe, a force is maintained constant between the probe and the sample. The SThM probe consists of a Pd resistive element deposited on a triangular-shaped silicon nitride tip. The change of resistance of this element with temperature, once calibrated, allows for the reading of the probe's temperature. The typical lateral resolution for such probe is between 10 and 50 nm\cite{spiece2018improving,spiece2021quantifying}. The resistive element of the probe is part of a Wheatstone bridge (see Figure \ref{fig1}c) and its resistance value is measured at 91 kHz \textit{via} a first lock-in amplifier. This electrical resistance is converted into a SThM probe temperature using the calibration procedure as explained elsewhere\cite{spiece2018improving,spiece2021quantifying}.

The back gate of the sample is modulated at the frequency $f$. This modulation and the SThM detection share the same electrical ground. Therefore, an electric field is applied between the tip and the sample. As described elsewhere\cite{harzheim2018geometrically,menges2016nanoscale}, we then demodulate the SThM probe temperature at the excitation frequency $f$ and its higher harmonics, a procedure which is crucial to decouple electrocaloric and other heating effects\cite{iguchi2023direct}.

This lock-in detection scheme drastically increases the signal-to-noise ratio of the temperature read-out, which is very useful when studying non-steady-state effects like the ECE, as the heat that is generated during each electric polarization/de-polarization cycle vanish rapidly in the sample and can be difficult to detect using conventional DC thermometry.

\subsection{Combined PFM and SThM}

The proposed measurement method relies on the AC modulation of the electric field between the probe and the sample. It is therefore necessary to demonstrate that, given the SThM probe geometry, an effective field modulation exists. We thus performed finite element modelling and piezoresponse force microscopy (PFM) measurements. PFM detects oscillations in the vertical or horizontal laser deflection in response to the converse piezoelectric effect in a material under an AC field modulation\cite{kalinin2004nanoelectromechanics,canetta2023quantifying}. While for PFM measurements an electrically-coated AFM tip is usually employed, here we use the SThM tip and rely on the Pd resistive thermometer to act as a ground while we apply a modulation on the back gate. In the following, we first present our PFM results and then perform finite element modelling of the tip-sample system. 

\begin{figure}[h]
  \includegraphics[width=\linewidth]{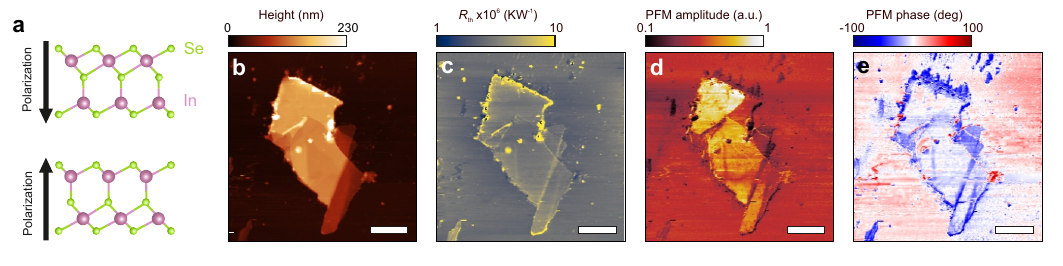}
  \caption{\textbf{Combined PFM and SThM on \inse.} (a) Crystal structure of \inse. The main polarization axis arise from the displacement of the middle Se atom. (b) Topography of the In${}_2$Se${}_3$ flake on the SiO${}_2$ substrate (c) SThM thermal conductance map. (d,e) Piezoresponse amplitude and phase acquired simultaneously. Scale bars: 4~$\mu$m.}
  \label{fig2}
\end{figure}

To test our method, we choose \inse as model system.  Several studies reported ferroelectric order due to the displacement of the middle Se atom\cite{xie2021ferroelectric,he2023epitaxial} (see \textbf{Figure \ref{fig2}a}). We obtain few-layer \inse flakes by using standard mechanical exfoliation of bulk crystals (HQ graphene) onto a doped Si wafer with 285 nm thermal oxide. \hl{We performed Raman spectroscopy on isolated flakes (see Supporting Information)}. Figure \ref{fig2}b shows a topography image of an isolated \inse flake. Ferroelectric materials also exhibit piezoelectric effects than can be probed with PFM. To increase the signal-to-noise ratio, we worked at the first contact resonant frequency which in our case was around $f = 125$ kHz (see Methods). The PFM amplitude and phase are shown in Figure \ref{fig2}d and e. Here, the vertical laser deflection signal was demodulated but in principle, the lateral signal can also be used. In Figure \ref{fig2}c, we show the SThM signal measured simultaneously with PFM, which demonstrates that the probe temperature and thus the thermal properties of the sample can be recorded at the same time as electrical fields are applied to the sample. This technique opens up new ways of investigating correlated thermal and electromechanical effects, such as polarization-dependent thermal conductivity, heat transport through domain walls or the interaction between temperature and polarization.

It is worth highlighting that PFM measurements are prone to artefacts, e.g.to short range and long range electrostatic interactions between the probe and the sample. Therefore, caution must be taken when interpreting PFM measurements obtained with this combined SThM-PFM measurement scheme. \hl{We also add that our results here are not a proof of ferroelectricity in \inse but a demonstration of the simultaneous tip-sample field modulation and probe temperature measurements.}

\begin{figure}[h]
\centering
  \includegraphics[width=0.85\linewidth]{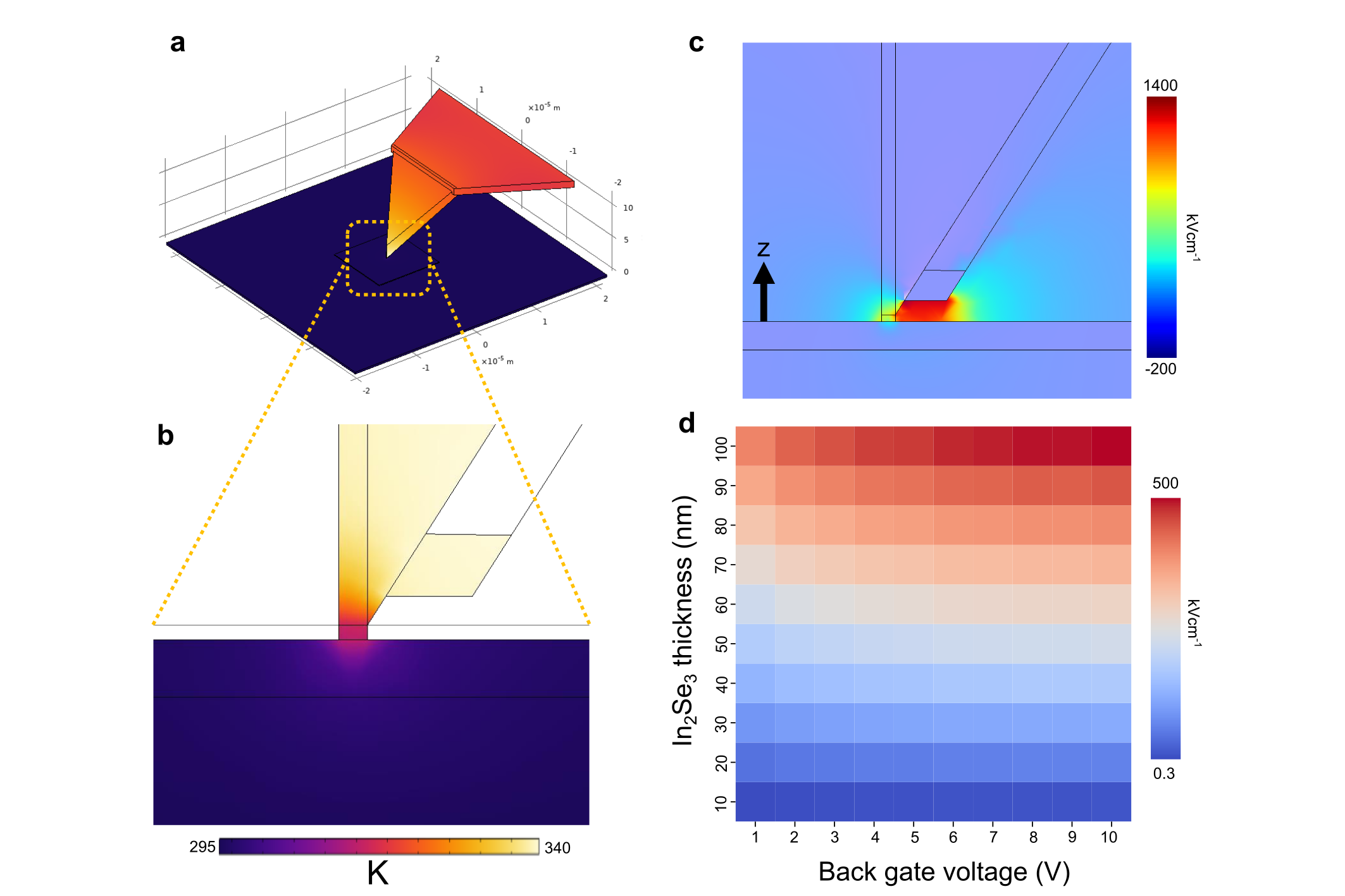}
  \caption{\textbf{Finite element modelling of the tip-sample system.} (a) Temperature distribution in the Joule heated probe (b) heat transfer between the tip and the sample. (c) Electric field strength distribution in the z-direction at the tip-sample contact for a 100 nm flake and 10 V on the back gate (d) Electric field strength as a function of back gate voltage and \inse thickness.}
  \label{fig3}
\end{figure}

To estimate the electric field strength on the sample we performed finite element simulations. Results are shown in \textbf{Figure \ref{fig3}}. We modelled the whole SThM probe undergoing Joule heating \textit{via} its Pd heater line (Figure \ref{fig3}a). The probe is in contact with a sample and the resulting temperature distribution of the system is shown in Figure \ref{fig3}b. As explained elsewhere\cite{spiece2018improving,evangeli2019nanoscale}, we developed a model to include the tip geometry in the measurements of the probe temperature. The sensing element in this probe geometry is distributed along the triangular tip. Most of the temperature drop occurs at the end of the tip while the rest of the Pd line remains at a constant temperature. 

In the same model, we implemented an electric field distribution calculation. As shown in Figure \ref{fig3}c, the field concentrates under the tip and mostly in the out-of-plane direction. The electric field strength at the tip-sample contact can be related to the gate voltage and flake thickness as shown in Figure \ref{fig3}d. We used $\epsilon_\mathrm{In2Se3} = 17$ \cite{Wu2015} for the relative permittivity of \inse. For 10 V back gate amplitude and 100 nm \inse thickness, we obtain 500 kVcm${}^{-1}$ which is beyond the coercive field required to switch polarization in \inse\cite{wan2018room,huang2022two}.

\subsection{Local ECE measurements: Scanning Electrocaloric Thermometry (SEcT)}

To drive electrocaloric effects, two AC excitation schemes were reported, the unipolar and the bipolar filed modulation. In the unipolar case, electric fields with only one field direction are applied to the sample which alternate between 0 and $E_\mathrm{max}$ (leading to only positive or only negative polarization in the material), while in the bipolar case, the field alternates between$-E_\mathrm{max}$ and $+E_\mathrm{max}$. For bipolar field modulation at the frequency $f$, the polarization follows the field, but the configurational entropy linked to the dipole orientation varies with $2f$ since the relaxation happens twice per cycle. Thus, the ECE also happens at $2f$.
In the case of a unipolar modulation, the oscillating electric field at $f$ creates an entropy variation at $1f$ and the electrocaloric effect also occurs at $1f$. However, in this case, the electrocaloric temperature variation $\Delta T$ is proportional to $\delta P/\delta E = P_r \delta P_\%/\delta E$ with the remnant polarization $P_r$ and $\delta P_\%/\delta E$ is the slope of the polarization change. If this slope is small, the temperature variation will also be small. \hl{Based on the time-dependent heat equation, we derived an analytical model of the frequency dependence for both unipolar and bipolar electrocaloric signal (see Supporting Information). In the bipolar case, no signal is expected at the 1f and the 2f signal should scale quadratically with the field amplitude. For the unipolar modulation, the signal occurs in both 1f and 2f channels and also scale quadratically.}

\begin{figure}[h]
\includegraphics[width=\linewidth]{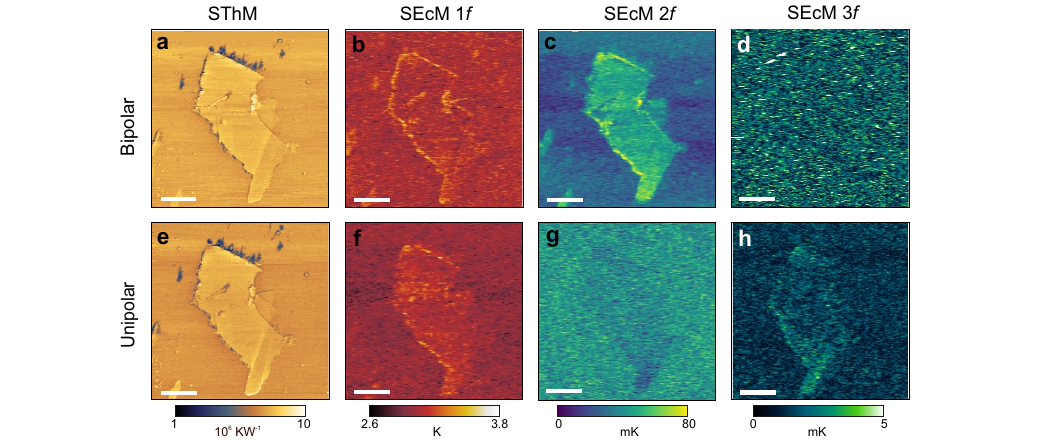}
  \caption{\textbf{Electrocaloric maps of \inse.} (a)-(d) Bipolar and (e)-(h) unipolar measurements. For both measurements, SThM, $1f$, $2f$ and $3f$ signals are displayed. Scale bars: 4 $\mu$m.}
  \label{fig4}
\end{figure}

In \textbf{Figure \ref{fig4}}, we show the thermal resistance and the local electrocaloric maps on an \inse flake using both unipolar and bipolar modulation. The modulation frequency of the electric field was $f= 537$ Hz and the field was modulated between 0 and $\pm E_{max} = 500$ kVcm${}^{-1}$ in the unipolar case and between $\pm E_{max} = 500$ kVcm${}^{-1}$ in the bipolar case. We simultaneously recorded the thermal resistance map and the $1f$, $2f$ and $3f$ responses. \hl{In the Supporting Information, we also show results on a hexagonal boron nitride flake where no signal is expected, as this material is only minutely polarizable.}

In Figure \ref{fig4}a and e, we observe a larger SThM signal on the \inse flake than on the underlying substrate. We attribute this contrast to different tip-sample contact and thermal interface resistance between the \inse flake and its substrate, as reported elsewhere\cite{evangeli2019nanoscale,spiece2022low}.

In the bipolar case, we observe a strong signal mainly in the $2f$ channel. This is expected as explained earlier by the dipole entropy variation also occurring at twice the modulation frequency. In the $1f$ response, we only observe the edges of the flake while the signal inside the flake and on the substrate have similar magnitude. We measured an average amplitude of $25\pm 6$ mK on the \inse flake. 

In the unipolar case, the signal appears stronger in the $1f$ response as expected. However, as explained above, the entropy variation is small as the polarization only varies from a remnant polarization state to a full polarization state. In the $2f$ channel, we can distinguish the flake appearing with a lower signal than the surrounding substrate. This reduction could be explained by the lower dielectric losses between the bare silicon oxide and the \inse of flake we develop below. Finally, we observe a small response in the $3f$ signal, too. The polarization itself is not necessarily linear with field and part of the electrocaloric signal can leak into other harmonics signals. In fast, it is possible to write $P(E) = \chi_1 E + \chi_2 E^2 + \chi^3 E^3$, where $\chi_n$ is the n-th order susceptibility. Thus, part of the ECE signal can occur at the 3rd harmonic as reported by Iguchi \textit{et al.}\cite{iguchi2023direct}. 

When applying an AC field to a dielectric material between two electrodes, several phenomena can generate heat in the process. First, in case of electrical leakage currents between the electrodes, Joule heating could contribute and also occur at the $2f$ frequency. However, for our 285 nm thick SiO$_x$ gate dielectric, we find an electrical resistance of $\gg 10$G$\Omega$  which would result in $\ll 10$nW power dissipation for an amplitude of 10V. Thus, the local heating from Joule effect can be safely neglected in our samples. Second, dielectric losses can be present in our system and can also manifest in the $2f$ signal. The power $P_C$ (equation \ref{PowC}) dissipated in the dielectric adds a constant background of $9\pm1$ mK to the map in Figure \ref{fig4}d. A slight change in signal contrast is expected on the \inse flake since the total capacitance of the system changes. Using $\epsilon_\mathrm{SiO2} = 3.9$, $\epsilon_\mathrm{In2Se3} = 17$ \cite{Wu2015}, $d_\mathrm{SiO2} = 285$ nm and $d_\mathrm{In2Se3} = 100$ nm we estimate a reduction of capacitance by 12\% in the \inse flake which i) cannot explain the contrast observed in Figure \ref{fig4}d and ii) which leads to an \textit{under}estimation of the electrocaloric coefficient using our method as the reduced capacitance leads to reduced heat generation through the dielectric as indicated in Equation \ref{PowC}. Finally, in ferroelectric materials with finite electric coercivity, a change in polarization with frequency $f$ would lead to hysteretic power losses $P_\mathrm{hyst} \propto A_\mathrm{hyst}f$, where $A_\mathrm{hyst}$ is the area of the hysteresis loop. This, however, leads to a constant (DC) heating of the sample which would not be visible in our AC experimental scheme and doesn't appear as a temperature drift in the SThM signal in Figure \ref{fig4}a.

\subsection{Frequency and field dependence}

Finally, we explore the frequency and field dependence of the electrocaloric effect. In this section, we focused on a bipolar modulation of the electric field to increase the signal. The frequency dependence is important to estimate if adiabatic temperature rise is reached. In the adiabatic condition, the heat generated in the ECE layer overcomes the heat dissipation to the environment. Thus, the adiabatic temperature rise enables an accurate estimation of the ECE effect and this makes comparison with other materials. 

\begin{figure}[h]
\centering
\includegraphics[width=0.5\linewidth]{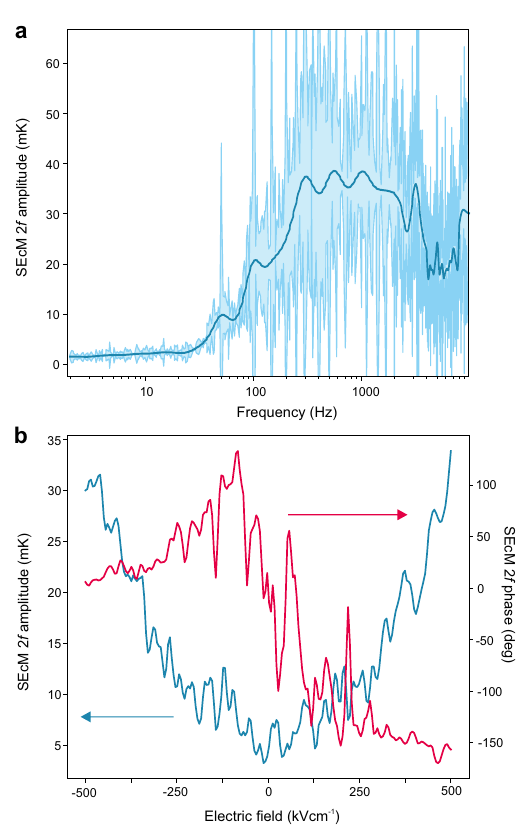}
  \caption{a) Frequency-dependence of the SEcT $2f$ amplitude signal measured at a field of $\pm$ 500~K~cm~kV${}^{-1}$. We added a smoothed guideline for the eye to improve the readability. b) Electric field amplitude dependence of the $2f$ amplitude and phase signal.}
  \label{fig5}
\end{figure}

In \textbf{Figure \ref{fig5}a}, we show the frequency dependence of the $2f$ electrocaloric effect. We obtained an increasing signal with increasing frequency. This is expected from a heat transport point of view. Electrocaloric heat is generated at each polarization switch cycle in the \inse flake. If the frequency is fast enough, this heat will warm up/cool down the layer before dissipating to the environment. On the measurement side, our technique is limited by the thermalization time of the SThM sensor. For this particular type of probe, several studies reported values from few tens of $\mu$s to ms in air\cite{puyoo2010thermal,puyoo2011scanning,gomes2015scanning,tovee2012nanoscale} which makes measurements above 10 kHz difficult, especially for higher harmonics. 

Figure \ref{fig5}b shows the $2f$ signal amplitude and phase as a function of the electric field $\pm E_{max}$. The field dependence allows to compute the electrocaloric strength and to measure the overall behaviour of the ECE effect. The signal amplitude increases with increasing field as expected. Indeed, a greater electric field variations allows for stronger polarization change in the material, resulting in more significant entropy change. The increase in amplitude is slightly non-linear. Non-linearities in the ECE temperature can arise from the polarization mechanisms inside the material, leading to non-linear entropy variations.   
This measurement enables us to compute the electrocaloric strength $\Delta T_{ECE}/\Delta E$ for \inse. We obtained $5 \times 10^{-5}$~K~cm~kV${}^{-1}$.

\section{Conclusion}
In this paper, we developed a new scanning probe microscopy method, Scanning Electrocaloric Thermometry (SEcT), and show that it can be used to quantify the local electrocaloric effect in a 2D ferroelectric material. The electric field modulation applied between the probe and the sample, coupled with lock-in detection, enabled the identification of ECE signals under both bipolar and unipolar modulation schemes. The AC measurement scheme drastically improved the signal-to-noise ratio and allowed us to disentangle other components from the ECE signal. We applied the method to \inse flakes and revealed distinct electrocaloric signal on the material with high spatial resolution, and extracted an electrocaloric strength parameter of $5 \times 10^{-5}$~K~cm~kV${}^{-1}$ for \inse.
Noteworthy, this study also introduced the combination of simultaneous PFM and SThM measurements opening new possibilities for electrothermal measurements at the nanoscale.

Because SEcT does not require the fabrication of complicated devices and can be readily applied to any meso- or nano-scale systems, it represents a powerful tool for investigating electrocaloric effects in a wide range of materials, including (multiferroic) van der Waals materials and their heterostructures. Additionnaly, SEcT can be used to study the impact of nanoscale defects, domain structures, or polarization landscape on the local electrocaloric response, providing valuable insights into material properties that are otherwise difficult to access with traditional methods. Future studies may also focus on optimizing material compositions and engineering nanoscale interfaces to enhance electrocaloric performance for applications in energy-efficient cooling technologies and compact thermal devices.

\section{Experimental Section}
\subsection{Scanning probe microscopy and thermometry}

This work was performed on a modified Thorlabs Atomic Force Microscopy (EDU-AFM1, Thorlabs Inc.) transformed into a Scanning Thermal Microscope. The tip (Kelvin Nanotechnologies Ltd.) consists in a grooved silicon nitride cantilever with a triangular tip. On the cantilever, two gold lines were deposited and on the tip a Pd line creates the sensing element. The tip is connected to a home-made Wheatstone bridge\cite{spiece2019quantitative}. The probe temperature is measured with a 91 kHz voltage on the bridge and demodulated with a lock-in amplifier (MFLI, Zurich Instruments Ltd.). The typical voltage drop on the tip is on the order of 0.3 V. 
The back gate of the sample is electrically connected with silver paste and wired to a waveform generator (DG900, Rigol inc.). The temperature of the probe is demodulated at the generator frequency and its harmonics.   
To record the various signals, we use the data acquisition capabilities of the lock-in software synchronised with the AFM scan.

In this study, we applied the principles of AC-SThM thermometry as detailed elsewhere\cite{harzheim2018geometrically,menges2016nanoscale}. However, the probe is different from the probe used in those studies and the sensor is spatially distributed\cite{spiece2018improving}. This must be taken into account to accurately relate the sample temperature variation and the global change of the SThM probe resistance.

In Spièce \textit{et al.}, we developed a model including the probe geometry and relating the average probe temperature $T_{av}$ to the system parameters: 
\begin{equation}
    T_{av} = \alpha Q_h + \beta T_m + \delta T_{air} + \gamma T_s
\end{equation}
where $\alpha, \beta, \gamma$ and $\delta$ are geometry and thermal resistance factors and $Q_h, T_m, T_{air}$ and $T_s$ are the heater power, microscope probe base temperature, air temperature and sample temperature, respectively. 
If we assume that during the AC temperature variation of the sample, only $T_s$ is varying, we can write
\begin{equation}
    \Delta T_{s} = \Delta T_{av} / \gamma
\end{equation}
$\Delta T_s$ is the sample temperature variation. It includes thermal resistance and geometry in the $1/\gamma$ factor.

\subsection{Contact resonance in PFM mode}

\begin{figure}
    \centering
    \includegraphics[width=0.4\linewidth]{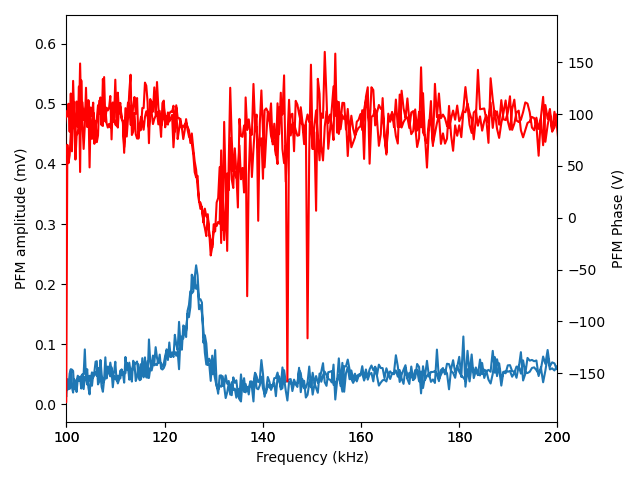}
    \caption{Frequency sweep of the amplitude and phase of the vertical PFM signal.}
    \label{fig:contactres}
\end{figure}

To perform the PFM measurements reported in the main text, we modulated the back gate voltage at the first contact resonance of the probe. In \textbf{Figure \ref{fig:contactres}}, we plot the frequency sweep of the PFM amplitude and phase where the contact resonance can be observed. 

\medskip
\textbf{Supporting Information} \par 
Supporting Information is available from the Wiley Online Library or from the author.

\medskip
\textbf{Acknowledgements} \par 
The authors acknowledge financial support from the F.R.S.-FNRS of Belgium (FNRS-CQ-1.C044.21-SMARD, FNRS-CDR-J.0068.21-SMARD, FNRS-MIS-F.4523.22-TopoBrain, FNRS-PDR-T.0128.24-ART-MULTI, FNRS-CR-1.B.463.22-MouleFrits, FNRS-FRIA-1.E092.23-TOTEM), from the EU (ERC-StG-10104144-MOUNTAIN), from the Federation Wallonie-Bruxelles through the ARC Grant No. 21/26-116, and from the FWO and FRS-FNRS under the Excellence of Science (EOS) programme (40007563-CONNECT).

\medskip

\textbf{References}\\
\bibliography{main}

\end{document}